# Business Type Classification via E-commerce Stage Model in Oil Industry in Iran

Mohammad Nassiry
Faculty of Information Science & Technology
University Kebangsaan Malaysia
Tehran, Iran
mohammad_nassiry@yahoo.com

Prof. Dr. Muriati Mukhtar
Faculty of Information Science & Technology
University Kebangsaan Malaysia
Kuala Lumpur, Malaysia
mm@ftsm.ukm.my

*Abstract*— since the strategies and plans for e-commerce development are different for different industries and since the oil industry is one of the most important industries in Iran, the scope of this research is thus confined to that of the oil industry in Iran. The main aim of this study is to identify and classify the different features of e-commerce development stages and features based on the different business types present in companies in the oil industry in Iran. In order to achieve both of these objectives a questionnaire was developed and administered online.  The questionnaire was distributed to forty representatives working in different companies. The collected data was classified and sorted and the priority e-commerce features was classified and displayed as triangles for each business type. Furthermore, the experts were asked to indicate the features which they implemented in their companies in order to know the most used features in each stage.  The results of this study give an insight to the practice of e-commerce for Iranian oil companies and can be used to strategize future directions for the industry in terms of e-commerce.

## I. INTRODUCTION

The globalization of markets and ICT, specially internet technology, creates a cost effective platform for companies to communicate and conduct commerce. By the invention of computing technology and communication systems, E-commerce covers the business activities between company and customer via electronic media (Stair and Reynolds, 2008). Therefore, companies have invested heavily in Information Technology, mainly for making automatic external and internal processes, and communication such as payroll, accounting, financial, human resources, and production. The significance of electronic commerce and its impact on reducing costs and increasing income has made researchers give serious attention to electronic commerce in the past few years.

Oil, gas and petrochemical industries have a significant role in the world energy market. Information Technology had a strong effect on the oil industry in many ways and takes the benefits of e-commerce. IT infrastructure and Internet can support the exchange of information between the segments of the oil industry.

Therefore the researcher decided to conduct this study to organize and classify the features of e-commerce development stages based on their priorities for helping oil industry to handle their business activities in the most effective way.

## II. LITRUTURE REVIEW

The literature review is focusing on three major areas. The first part explained E-commerce adoption models and e-commerce adoption in Iran. The second part of literature addresses development stages of e-commerce, a stage model and its features. The third area shows e-customer Chain Model in Oil Industry and classification of e-business items in oil industry.

### A. E-commerce Adoption Model

E-commerce adoption could be examined and evaluated at any levels for the e-commerce successful implementation. Some of the frameworks and models like, Mosaic (1997), Mcconnell (2000), APEC (2000) and Harvard model (Kirkman et al., 2002), study of e-readiness at national and macro-level. Meanwhile, some of the e-commerce adoption variables refers mainly at the national or macro level is national or macro level is also referred in micro models like, Rashid and Qirim framework (Rashid & Qirim, 2001), the Ling model (Ling, 2001) and Wang and Tsai (Wang & Tasi, 2002). Therefore, e-commerce adoption might eventually be split on two distinct levels: (1) National level, and (2) Organizational level. (Elahi, Hassanzadeh, 2010), "A framework for evaluating electronic commerce adoption in Iranian companies", this model conducting the e-commerce is an extension to Tan et al (2007) and stresses on organizational or micro level e-commerce while some models emphasize on studying e-commerce at the macro or national level. Figure 2.1 shows a framework for evaluating electronic commerce adoption in Iranian companies.

Figure 2.1 a framework for evaluating electronic commerce adoption in Iranian companies

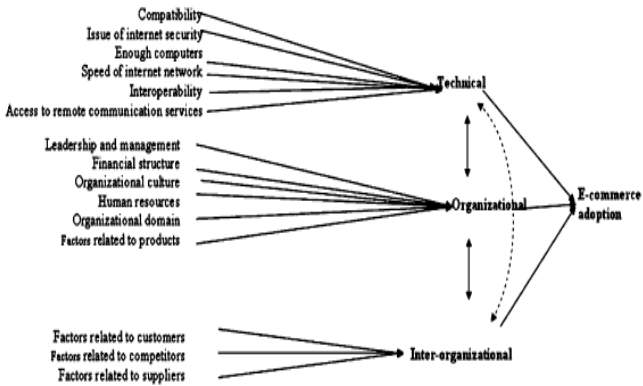

According to Wu (2004) and Zulkifli (2001), the e-commerce adoption process starts with knowledge mind; continued by attitude training, decision-making and implementation. According to Rao, Metts, and Monge (2003) electronic commerce development and its implementation is taking place in 4 stages (1) Presence: at this stage most businesses making their first steps in e-commerce by showing their business brochures and products offering on the web; (2) Portals: at this stage the portals is regarded as the introduction of two-way communication, client or vendor order placement, the use of profiling and cookies; (3)Transaction integration: this stage (TI) is distinguished from the portals stage primarily by the availability of financial transactions among partners; and (4) Enterprises integration: enterprises integration (EI) relate to full integration of business processes to the degree, which old-line business is identical from online business.

### B. E-commerce Stage Model and their Features

A stage model is a set of descriptors, which characterize the evolving nature of electronic commerce. Such descriptors are for example, brochure, in-line catalogs, contact, one-way and 2-way communication, connection information, on-line banking transactions, etc.

Regarding the e-commerce adoption levels, there are various models of e-commerce development offered by a number of researchers, for instance, O'Connor and O'Keefe (1997) and Timmers (1999) explains the e-commerce business model. The first writers characterized the models by the degree of operations, transactions and the degree of content. Timmers (1999) featuring a business model by 2 dimensions: degree of innovation and degree of functional integration. The models explain how a company is using e-commerce to run their business "The E-commerce Maturity Model" (KPMG, 1997), "Stages of e-commerce development" (Rao, Metts, and Monge, 2003), "Ecommerce Adoption Model" (Daniel, Wilson and Myers, 2002), and "the Stages of Growth for E-Business" (SOGe) model (Prananto, McKay, and Marshall, 2003); each model with particular conditions and features in order to explain the levels of e-commerce adoption by enterprises. However, all of them agreed that a higher level would need to applications that are more complex and would produce more advantages than the previous stages.

According to Rao, Metts, and Monge, (2003), there are 4 stages of e-commerce development using that are the presence stage, portal stages, transaction integration stage and enterprises integration stages. Every stage has various features. Figure 2.2 shows e-commerce development stage model.

Figure 2.2 shows e-commerce development stage model

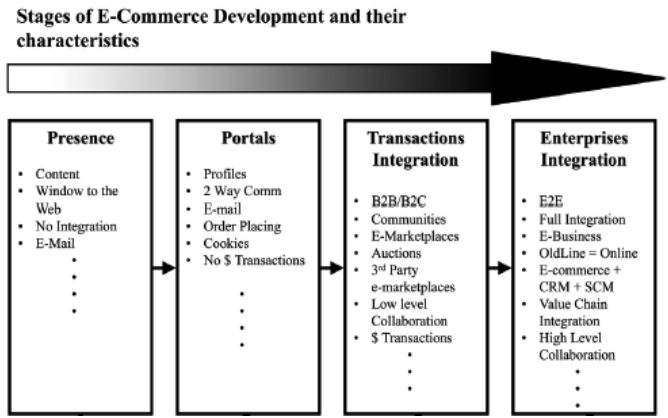

#### B.a. Presence Stage and its Features

Most businesses make their initial steps in electronic commerce by showing their corporate brochure and goods offering on a website (Timmers, 2000). The presence stage includes the first steps that businesses do to engage in the digital environment. A company should have a product which introduces itself and via this medium and has taken the necessary steps to ensure that the website is appealing and user friendly. Table 2.1 shows the features of Presence stage for e-commerce development stage that has been pointed by different scholar.

Table 2.1 the features of Presence stage for e-commerce development stage

| No | Features | Sources |
|----|----------|---------|
| 1 | Content | Jeffacoate, 2000 |
| 2 | Windows to the web | Barry, 2000 |
| 3 | No Integration | O'Conner and O'Keefe, 1997 |
| 4 | E-Mail | Timmers, 2000 |
| 5 | Commitment | Roa. 2003 |

*B.b. Portals Stage and its Features*

The portal stage is considered the introduction of two-way communication, client or vendor order placement, the use of cookies and profiles. The major difference between this stage and the presence stage is the ability of two-way communications between the company and clients (B2C) and between companies (B2B). The data in the presence stage may be combined with facilities for ordering, product information, and goods and quality surveys. This enables not only the attraction of new clients, but also enables the company to attract and keep visitors, and to relate them to their personal preferences for personalization purposes (Le and Koh, 2001). Another benefit of this stage is the ability to link information displayed through inventory information, and search features to users (Timmers, 1999). Table 2.2 shows the features of Portals stage for e-commerce development stage that has been pointed by different scholar.

Table 2.2 the features of Portals stage for e-commerce development stage

| No | Features | Sources |
|----|----------|---------|
| 1 | Profile | Le and Koh, 2001 |
| 2 | 2-way communication: | Chapman, 2000 |
| 3 | Order placing | Roa, 2003 |
| 4 | Cookies | Koh, 2001 |
| 5 | No money transaction | Roa and Metts. 2003 |
| 6 | Usability | Le and Koh, 2001 |
| 7 | E-mail | Monge 2003 |
| 8 | Culture and language | Zhivago, 2000 |

*B.c. Transaction Integration Stage and its Features*

The transactions integration stage is separate from the portals stage primarily by the presence of the financial transactions between the partners. This in turn required greater technical capacity, and therefore, the companies would face new challenges of overcoming.

An organization at this stage should have a higher level of technical capacity in order to perform the E-commerce business (Chesher and Skok, 2000). Table 2.3 shows the features of Transaction integration stage for e-commerce development stage that has been pointed by different scholar.

Table 2.3 the features of Transaction integration stage for e-commerce development stage

| No | Features | Sources |
|----|----------|---------|
| 1 | Communities | Timmers, 2000 |
| 2 | 3rd party marketplace | Timmers, 2000 |
| 3 | Auction | Bishop, 1999 |
| 4 | Money transactions | Frakas-Conn, 1999 |
| 5 | Low level collaboration | Chesher & Skok, 2000 |
| 6 | Integration | Chesher & Skok, 2000 |
| 7 | Competitive payment system | Fariselli, 1999 |
| 8 | E-marketplace | Timmers, 2000 |
| 9 | B2B/B2C | Timmers, 2000 |
| 10 | Security and privacy | USSBA, 2000, Timmers, 2000, Bollo and Stumm, 1998 |

*B.d. Enterprises Integration Stage and its Features*

Enterprises integration is related to the full integration of the business processes to the extent that the old-line business is different from online business. This level of integration requires high levels of cooperation between clients and suppliers. Enterprises integration provides full integration of B2B and B2C business, including the value chain integration. This level of integration uses the E-commerce systems to managing the supply chain (SCM) and customer relationships (CRM). This level of integration is E-commerce + SCM + CRM. This stage is a bit of a perfect concept for the "E-global" environment. Many of the devices of this wise still have technical issues and over-whelming integration problems. Table 2.4 shows the features of Enterprises integration stage for e-commerce development stage that has been pointed by different scholar.

Table 2.4 the features of Enterprises integration stage for e-commerce development stage

| No | Features | Sources |
|----|----------|---------|
| 1 | Full integration | Roa & Mette, 2003 |
| 2 | E-business | Roa & Mette, 2003 |
| 3 | Old line = Online | Roa, Mette & Monge, 2003 |
| 4 | E-commerce +CRM+SCM | Lacerra et al, 1999, Krause et al, 1998 |

| 5 | Value chain integration | Roa & Mette, 2003 |
| 6 | High level collaboration | Krause et al, 1998 |
| 7 | E2E: End to End | Lacerra et al, 1999 |

*C. E-customer Chain Model in Oil Industry*

A key concept, which emphasizes the role of IT is customer chain (Wei, J, 2009). According to June Wei, 2009, electronic customer chain is defined as the use of firms IT infrastructure or Internet functions to share data with every member of the customer chain in order to ensure the improvement and synergies between the links in the chain.

The procedures within each stream (upstream, midstream, and downstream) are shown as electronic processes (e-processes) since every process should be carried out via electronic media (contains IT aid). The e-customer is the customer who is interacting with the oil and gas companies through electronic media.

The upstream operations for the exploration and production sectors which finding and producing petroleum and gas. The midstream segment shops, markets and transports goods including petroleum, gas and gas liquids. The downstream segment includes refining oil, petrochemical products, dealers, and petrol stations to distribute petroleum products to final e-consumer.

The three sectors (Downstream, Midstream, and Upstream) have a relationship as a customer or as a provider with the party that is next or preliminary to it in a supply chain. Some act as internal e-customers and certain external e-clients (Wei J, 2009). An external e-customer directly purchase the product or service from a company, such as the late e-consumer or in the case that an independent operator sells its product to a company in the midstream or downstream segment. Wei remains that an internal e-customer is a group in the distribution channel that gets processes, services and products from another in the organization. The various possible relationships can be classified in four e-business categories: business-to-business (B2B), consumer-to-business (C2B), business-to-consumer (B2C), and business-to-internal (B2I). Figure 2.3 shows an oil e-customer chain model development.

Figure 2.3 shows an oil e-customer chain model development

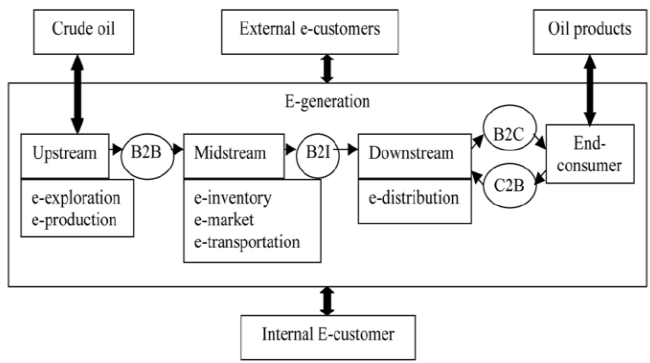

*D. Classification of E-business items in Oil Industry*

(Mihlmester and McKelvey, 2006) developed a set of e-business applications for the general energy services. (Ende and Wei 2007, pp. 489-501) developed a set of e-business applications in the oil industry. According to Mihlmester and McKelvey's e-business applications and the expansion of the security items developed by Ende and Wei, of e-business model in oil industry (Wei and Ende, 2009), thirty-six e-business solution products in the oil industry are mapped and obtained from the developed e-customer chain model in terms of four e-business category comprising: B2B, B2C, C2B, and B2I. Each category has its own application, which can be implemented in the organization to enhance business worth on physical or information-processing components.

Based on the study of existing e-business solution items to these dominant companies, an e-business solution adoption triangle is built. The solution items that equal or more than 80% of the firms implemented are categorized as most commonly used elements (Most Adopted Items), these equal or less than 25% least frequently used elements (Least Adopted Items), and the remaining items (Enhanced Items).

## III. RESEARCH METHODOLOGIES

The survey comes from the Roa research. The questionnaire was replicated in this research that has been designed based on the literatures "E-commerce development in SMEs, A stage model and its implications", and "Customer focused E-Business model for oil industry". The results of their studies show that the features and business types have been well identified and their model is a big and clarify the development stages of e-commerce in oil industry.

In this study, following the design of the questionnaire of the pilot test was carried out through e-mail by submitting this survey to a few experts in this area. They assist us with responding to questions and ask them to comment concerning the structure of the survey. This measure has set the validity of the content of our survey.

After this, the reliability of responses by applying Cronbach's alpha coefficient in SPSS was calculated.

Design Likert's scale questionnaire in connection with each development stage and its features and the responses will be based on each business types B2B, B2I, B2C and C2B and distribute online to IT experts in oil and petrochemical companies in Iran. Data collected from the questionnaire are analyzed and calculating composite score (mean, sum) to find importance of each features of e-commerce development stage from different business types in oil and petrochemical industries. Four different triangle s are designed for four existing business types based on the analysis of the data. Implemented of features stages of e-commerce development is defined based on the responds from the questionnaire.

IV. DATA ANALYSIS

Data gathered from the questionnaire were analysed by using SPSS software version 19 and descriptive analysis through frequency statistic. We follow by measures of central tendency and location (mean and sum) for taking the result of importance features stages of e-commerce development by different business types. Based on the analysis of the data the priorities of features stages of e-commerce development in the oil industry and the implementation stages of e-commerce development in Iranian oil companies in this study will be analysed and explained. The next step following the end of the information collection is to organize the information into a useful form so that the trend, if any, arising from the information can be displayed easily.

Based on the examination of existing features of stages of e-commerce development for these dominant companies, a triangle for priorities of features of e-commerce development stage in each business type is constructed. The features that equal to or more than 80 of the composite score (sum) are classified as most important features (80 above), those equal to or less than 40 least important features (40 below), and the rest is medium priority features. With this result, every company knows about its priorities in the various stages of its internal and external communications.

*A. Business Type Classification Triangle for Business to Business*

In figure 4.1, from 30 features of e-commerce development stage the 21 most important features are A1, A2 ,A4, A5, B1, B2, B3, B6, B7, C4, C5, C6, C7, C9, C10, D1, D2, D3, D4 and D5 (SUM= 80 or above), which are usually found in the Presence, Portals, Transaction integration and Enterprises integration stages categories. The 4 least important features stages of e-commerce development are A3, B8, C1 and C3 (SUM= 40 or below). These features are shown in the Presence, Portals and Transaction integration stages categories. The 5 of the medium priority features are B4, B5, C2, C8 and D6 which shown in the Portals, Transaction integration and Enterprises integration stages categories. Moreover, 6 of thirty features stages of e-commerce development are the features that are the most important with highest rated among the analyzed companies by respondent experts (A2, B1, B2, C10, D1 and D5). The least commonly features are C1 and C3. Figure 4.1 shows the priorities features of e-commerce development in B2B.

Figure 4.1 the priorities features of e-commerce development in B2B

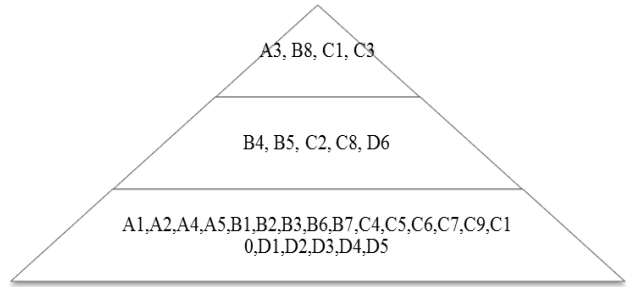

*B. Business Type Classification Triangle for Business to Internal*

In figure 4.2, from 30 features of e-commerce development stage the 21 most important features are A1, A2, A3, A4, A5, B1, B2, B3, B6, C5, C6, C7, C8, C9, C10, D1, D3, D4, D5, D6 and D7 (SUM= 80 or above), which are usually found in the Presence, Portals, Transaction integration and Enterprises integration stages categories. The 5 least important features stages of e-commerce development are B7, B8, C1, C2 and D2 (SUM= 40 or below). These features are shown in the Portals, Transaction integration and Enterprises integration stages categories. The 4 of the medium priority features are B4, B5, C3 and C4 which shown in the Portals and Transaction integration stages categories. Moreover, 5 of thirty features stages of e-commerce development are the features that are the most important with highest rated among the analyzed companies by respondent experts (A5, B2, C10, D4 and D6). The least commonly feature is B8. Figure 4.2 shows the priorities features of e-commerce development in B2I.

Figure 4.2 the priorities features of e-commerce development in B2I

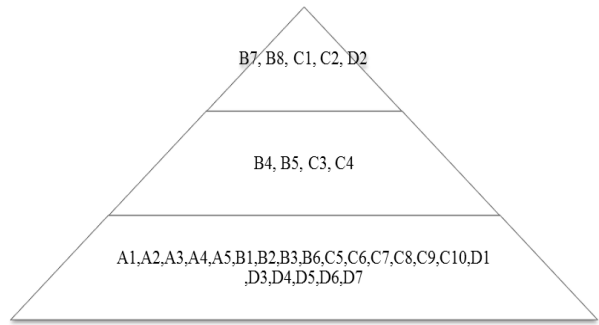

*C. Business Type Classification Triangle for Business to Customer*

In figure 4.3, from 30 features of e-commerce development stage the 12 most important features are A1, A2, A3, A4, A5, B1, B2, B6, C4, C6, C10 and D3 (SUM= 80 or above), which are usually found in the Presence, Portals, Transaction integration and Enterprises integration stages categories. The 8 least important features stages of e-commerce development are B8, C1, C2, C3, C5, D2, D6 and D7 (SUM= 40 or below). These features are shown in the Portals, Transaction integration and Enterprises integration stages categories. The 10 of the medium priority features are B3, B4, B5, B7, C7, C8, C9, D1, D4 and D5 which shown in the Portals, Transaction integration and Enterprises integration stages categories. Moreover, 4 of thirty features stages of e-commerce development are the features that are the most important with highest rated among the analyzed companies by respondent experts (A1, A5, B1 and D3). The least commonly features are B8 and C5. Figure 4.3 shows the priorities features of e-commerce development in B2C.

Figure 4.3 shows the priorities features of e-commerce development in B2C.

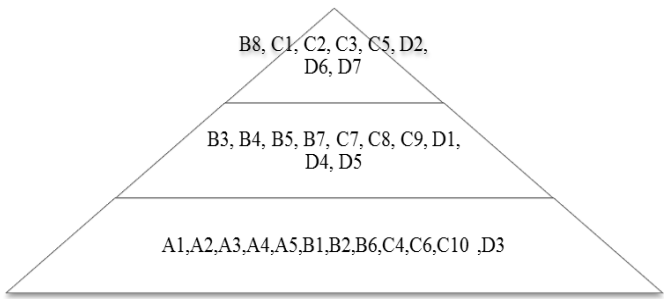

### D. Business Type Classification Triangle for Customer to Business

In figure 4.4, from 30 features of e-commerce development stages the 0 most important features. The 8 least important features stages of e-commerce development are A3, B5, B8, C5, C9, D2, D4 and D6 (SUM= 40 or below). These features are shown in the Presence, Portals, Transaction integration and Enterprises integration stages categories. The 22 of the medium priority features are A1, A2, A4, A5, B1, B2, B3, B4, B6, B7, C1, C2, C3, C4, C6, C7, C8, C10, D1, D3, D5 and D7 which shown in the Presence, Portals, Transaction integration and Enterprises integration stages categories. Moreover, 1 of thirty features stages of e-commerce development is the feature that is the highest rated among the analyzed companies by respondent experts (B6). The least commonly feature is B8. Figure 4.4 shows the priorities features of e-commerce development in C2B.

Figure 4.4 the priorities features of e-commerce development in C2B

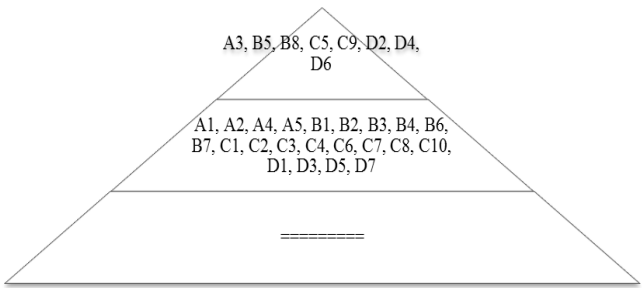

### E. IMPLEMENTED FEATURES

The flexibility of these 31 oil and petrochemical companies' stages of e-commerce development implemented status was studied in the present research. The total number of features and percentages for each stages of e-commerce development (Presence, Portal, Transactions integration and Enterprises integration) from each business types (B2B,B2I,B2C and C2B) are presented in ascending order in the following tables.

| Business type | Most implemented features In companies |
|---|---|
| B2B | A2, A4, A5, B1, B6, B7, C2, C4, C10, D2, D5 |
| B2I | A1, A4, B1, B2, B5, B7, C2, C6, C8, C10, D2, D3, D5 |
| B2C | A1, A2, A4, B1, B2, B4, B7, C2, C4, C6, C8, C10, D2, D3, D5 |
| C2B | A1, A2, A4, B1, B3, B4, B7, C2, C3, C4, C7, C8, C10, D2, D3, D6 |

## V. DISCUSSION AND CONCLUSION

The main results of this study are: a) discovering the priorities of features of e-commerce development stage in each business type. The outcome indicates that presence, portals and transactions integration are the most important stages for oil companies b) The most implemented features of e-commerce development stagein companies in each business type. The results from the current study indicate that the classification of features in the oil industry is critical to the success of the oil business activities. The main benefits of the results are presented below:

1. First, using this classification can simplify the flow of information, eliminate waste and reduce costs and times. So that the organization seeks satisfied during the oil and petrochemical industry. This business type classification through e-commerce stage model is used as a base for managers when considering business process redesign.

2. Second, it can make managers aware of the number of implemented features in e-commerce is not as important as implementing the most important features. Implementing the most important features can make the company e-commerce process faster and more efficient.

### A. Comparison of high priority features and implemented features

The comparison shows that although in B2B business type A1, B2, B3, C5, C6, C7, C9, D1, D3, D4 are high priority features, none of the companies are used these features. This issue is correct for the other business types. In B2I business type A2, A5, B3, B6, C5, C7, C9, D1, D3, D4 are high priority features, none of the companies are used these features. In B2C business type A3, A5, B6 are high priority features, none of the companies are used these features. In C2B business type A5, B2, B6, C1, C6, D1, D5, D7 are high priority features, none of the companies are used these features. Therefore, companies by implementing these features in their e-commerce system can increase their performance in customer relationship, document management, and procurement process.

ACKNOWLEDGMENT

Sincere gratitude is hereby extended to the following that never ceased in helping until this paper is structured.

I am grateful for the unwavering moral, emotional and financial support of the proponent's my supervisor Prof. Dr. Muriati Mukhtar, family and friends.

Above all, utmost appreciation to the almighty God for the divine intervention in this academic endeavor

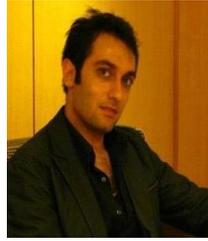
Mohammad Nassiry received his B.Sc. degree in IT in 2010 and he is doing his last semester in M.Sc. degree in MIS (Management Information System) in University Kebangsaan Malaysia. He is interested in E-Supply Chains, Business and system analysis, decision support system and Management of Information Technology
(E.mail: mnassiry62@gmail.com)

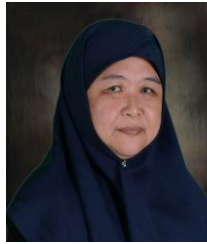
Assoc. Prof. Dr. Muriati Mukhtar is the head of research center for software technology and management (Softam) along with the head of service science research group in Faculty of Information Science & Technology in University Kebangsaan Malaysia. She obtained her PhD from UTMalaysia, SSn from UKMalaysia and BSc from UMIST. Her specialization areas are Service Science, E-Supply Chains and Simulation and Modeling. (E.mail: mm@ftsm.ukm.my)